\documentclass[a4paper]{article}
\pdfoutput=1
\usepackage{ISCSLP2021}
\usepackage{amsfonts} 
\usepackage{amsmath,graphicx,bm}
\usepackage{booktabs,multirow} 

\title{Mask Detection and Breath Monitoring from Speech: on Data Augmentation, Feature Representation and Modeling}

\name{Haiwei Wu$^{1}$, Lin Zhang$^{1}$, Lin Yang$^3$, Xuyang Wang$^3$, Junjie Wang$^3$, Dong Zhang$^4$, Ming Li$^{1,2}$}
\address{
    $^1$Data Science Research Center, Duke Kunshan University, Kunshan, China\\
    $^2$School of Computer Science, Wuhan University, Wuhan, China\\
    $^3$AI Lab of Lenovo Research, Beijing, China\\
    $^4$School of Electronics and Information Technology, Sun Yat-sen University, Guangzhou, China
    }
\email{ming.li369@dukekunshan.edu.cn}

\begin{document}

\maketitle

\begin{abstract}
This paper introduces our approaches for the \textit{Mask} and \textit{Breathing} Sub-Challenge in the Interspeech C{\scriptsize OM}P{\scriptsize AR}E Challenge 2020. For the mask detection task, we train deep convolutional neural networks with filter-bank energies, gender-aware features, and speaker-aware features. Support Vector Machines follows as the back-end classifiers for binary prediction on the extracted deep embeddings. Several data augmentation schemes are used to increase the quantity of training data and improve our models' robustness, including speed perturbation, SpecAugment, and random erasing. For the speech breath monitoring task, we investigate different bottleneck features based on the Bi-LSTM structure. Experimental results show that our proposed methods outperform the baselines and achieve 0.746 PCC and 78.8\% UAR on the Breathing and Mask evaluation set, respectively.
\end{abstract}

\section{Introduction}\label{introduction}

Besides linguistic information, speech delivers various kinds of paralinguistic information, incorporating language, accent, gender, channel, emotion, psychological states, etc. \cite{schuller2013paralinguistics}. To explore the automatic identification of paralinguistic attributes in audio data, the Interspeech COM{\scriptsize PUTATIONAL} PAR{\scriptsize ALINGUISTIC CHALLENG}E (C{\scriptsize OM}P{\scriptsize AR}E) is held at the Interspeech conference each year since 2009 \cite{schuller2011recognising}. In 2020, the twelfth C{\scriptsize OM}P{\scriptsize AR}E \cite{schuller2020interspeech} focuses on three sub-challenges: 
Elderly Emotion Sub-Challenges (ESC), Breathing Sub-Challenge (BSC), and Mask Sub-Challenge (MSC).  

C{\scriptsize OM}P{\scriptsize AR}E encourages participants to develop task-depen- dent/task-independent features and techniques in different tasks. In this year, for both the MSC and BSC task, the organizer provides three traditional feature sets: {\scriptsize OPEN} SMILE acoustic feature set \cite{Eyben2010openSMILE} (6373-dimensional for MSC, 130-dimensional for BSC), Bag-of-Audio-Words (BoAW) extracted by {\scriptsize OPEN}XBOW \cite{Schmitt2017openXBOW}, and {\scriptsize AU}DEEP \cite{Freitag2017auDeep}. Finally, a support vector machine (SVM) based classifier/regressor is employed. Besides the above task-independent features, for the MSC, the baseline system extracts a 2048-dimensional D{\scriptsize EEP} S{\scriptsize PECTRUM} feature \cite{Amiriparian2018Bag, Amiriparian2017Snore, Amiriparian2017Sentiment} from a pre-trained convolutional neural network, which achieves the best unweighted average recall (UAR) of 70.8\% among all single systems on the development set. For the BSC, a sequential regression problem, the baseline system provides an end-to-end deep sequence modeling approach (CNN-LSTM) \cite{Trigeorgis2016Adieu, tzirakis2018end2you}, achieving the highest Pearson’s Correlation Coefficient (PCC) of 0.731 on the development set.

In recent years, deep learning based methods have achieved state-of-the-art performances in many paralinguistic tasks \cite{Yeh2019, Amiriparian2017Snore, Danwei2017e2e, Tang2018e2e, Haiwei2019The}. Structures of the convolutional neural network (CNN) \cite{Amiriparian2017Snore, Danwei2017e2e, Tang2018e2e, Haiwei2019The} and long short term memory (LSTM) \cite{Yeh2019} is playing an increasingly important role in feature extraction and modeling. Thus, our works concentrate on these deep neural networks (DNN) based approaches.

Given that the BSC task is a sequence-to-sequence regression task, we pay attention to the high-performance Bi-LSTM network. Besides, we experiment with bottleneck features to investigate whether phonetic information is useful for predicting breathing states. For the MSC task, we implement two kinds of convolutional neural network systems to extract high-level embeddings from the filter-bank energies (Fbank). SVM is employed on the top of these embeddings to make decisions \cite{Haiwei2019The}. We also investigate three approaches of data augmentation, which achieve significant improvements in our end-to-end framework. 

The rest of this paper is organized as follows: tasks and databases are presented in section 2. Features and modeling are given in section 3, and experimental results are provided in Section 4. Section 5 concludes the work.

\section{Task description and database}

\begin{figure*}[htb]
\begin{minipage}[b]{.23\linewidth}
  \centering
  \centerline{\includegraphics[width=4.0cm]{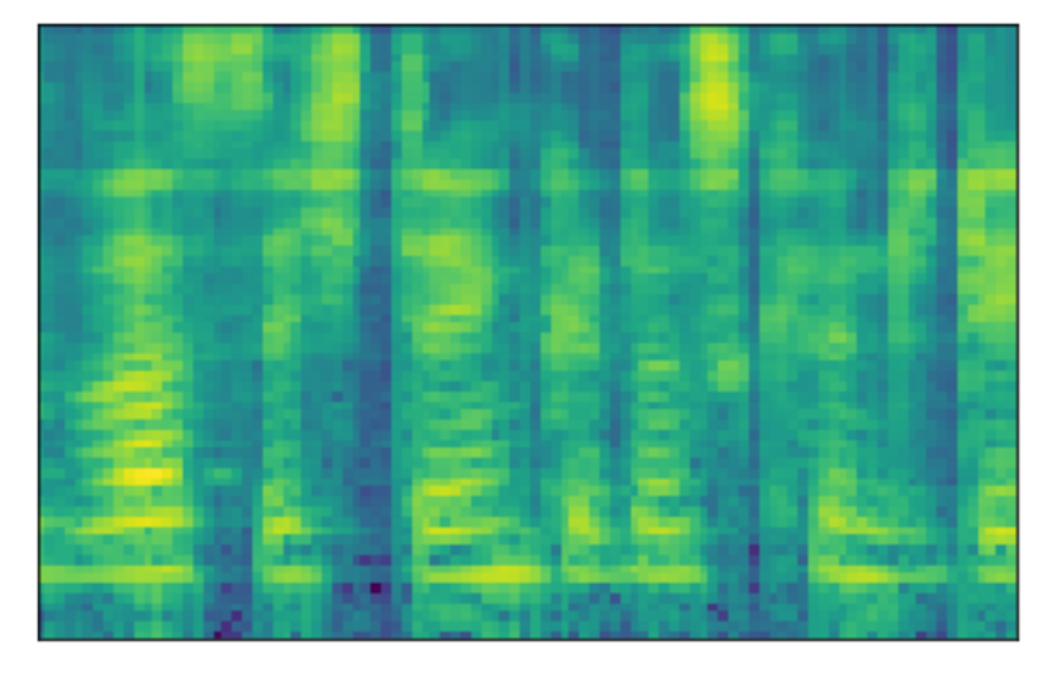}}
  \centerline{Original Fbank}\medskip
\end{minipage}
\hfill
\begin{minipage}[b]{0.23\linewidth}
  \centering
  \centerline{\includegraphics[width=4.0cm]{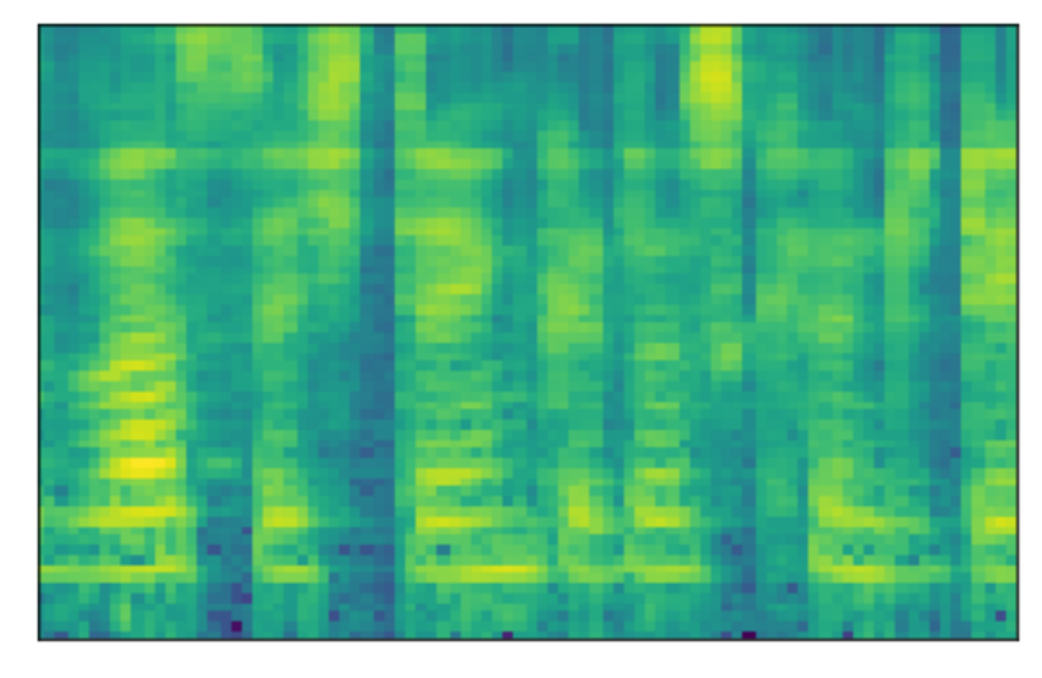}}
  \centerline{Speed pertubation}\medskip
\end{minipage}
\hfill
\begin{minipage}[b]{0.23\linewidth}
  \centering
  \centerline{\includegraphics[width=4.0cm]{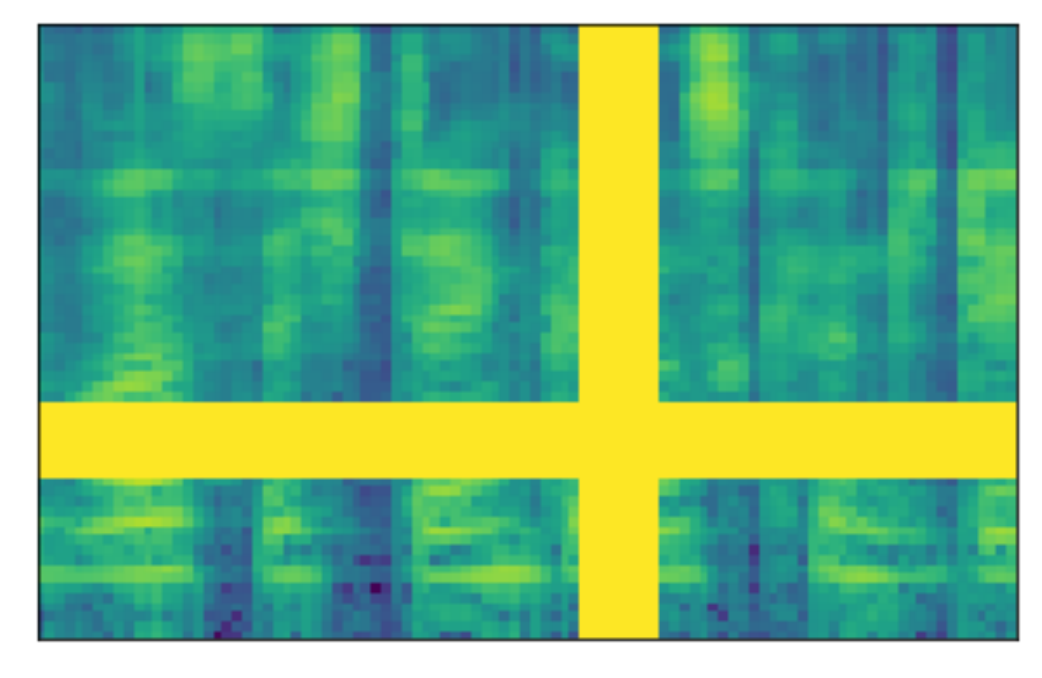}}
  \centerline{SpecAugemnt \cite{Park2019}}\medskip
\end{minipage}
\hfill
\begin{minipage}[b]{0.23\linewidth}
  \centering
  \centerline{\includegraphics[width=4.0cm]{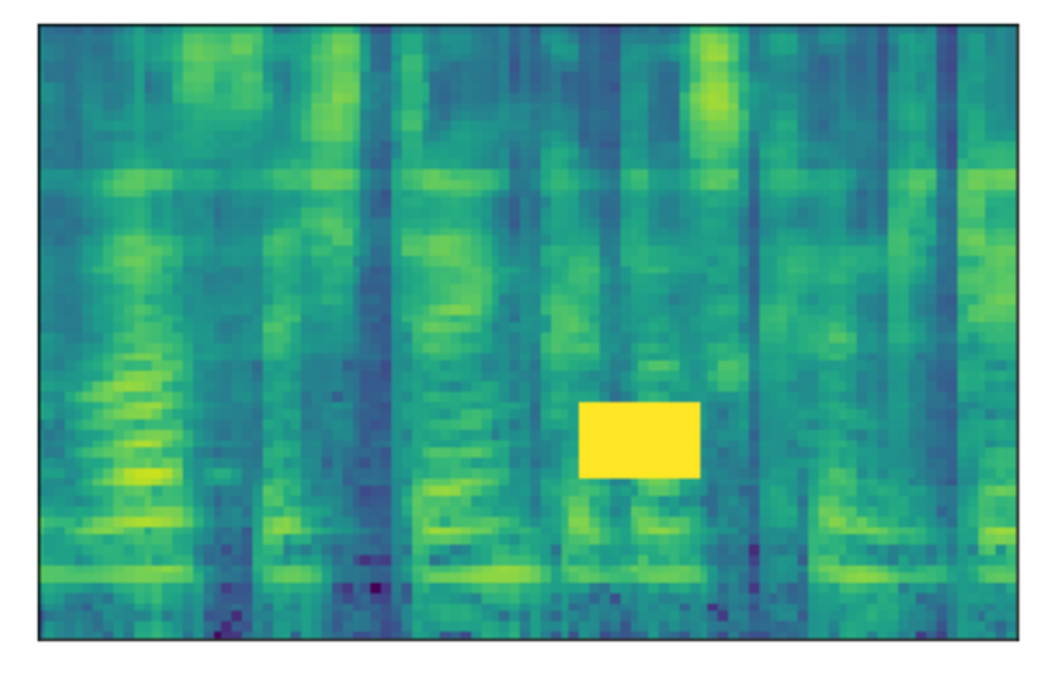}}
  \centerline{Random erasing \cite{zhong2020random}}\medskip
\end{minipage}
\vspace*{-0.5cm}
\caption{Effects of different data augmentation methods on the BSC}
\vspace*{-0.5cm}
\label{data_augmentation}
\end{figure*}

\subsection{Mask detection}\label{mask_task}

The Mask Sub-Challenge is a binary classification task to identify whether a speaker is wearing a facial protective mask. As COVID-19 spreading around the world, many people start to wear masks to protect themselves. Researchers found that wearing a mask affects speech production, resulted from muscle constriction, increased vocal effort, and transmission loss  \cite{llamas2009effects}. Experiments on speaker recognition \cite{Saeidi2016Analysis} and speech recognition \cite{Ravanelli2013Distant} shows that wearing masks would bring degradation to the performance of original speech systems, which indicates the necessity to detect whether the speaker is wearing a mask or not. The Mask Augsburg Speech Corpus (MASC) \cite{schuller2020interspeech} is used for the MSC task. In this database, the audio of 32 native German speakers is segmented into chunks of one-second duration for training and evaluation.

\subsection{Speech breath monitoring}

The Breathing Sub-Challenge is a sequential regression task that predicts a temporal breathing signal from recorded speech. Breathing condition is related to the speaker’s hesitation, duration, and emphasis of utterance \cite{Rochet2013The, Winkworth1994Variability}. Developing algorithms that could reliably monitor breathing states can provide vital information for doctors to make respiratory and speech planning and help singers to better manipulate their breath sound. For the BSC task, a subset of the UCL Speech Breath Monitoring (UCL-SBM) database \cite{schuller2020interspeech} is used. All 49 speakers reported English as a primary language covering a wide range of regional accents, sociolect, and ages. For each speaker, four minutes of spontaneous speech is recoded under a quiet office space.

\section{Features and modeling}

This section describes our feature extraction, data augmentation, and modeling of the MSC and BSC. For both tasks, we employ log-Fbank as our input acoustic features. For the MSC, gender-aware and speaker-aware features act as a complement to the log-Fbank. Speed perturbation, SpecAugment \cite{Park2019} and random erasing \cite{zhong2020random} are adopted for data augmentation. Deep convolutional networks are trained to extract embeddings, followed by a back-end SVM for classification. In the BSC task, we also explore bottleneck (BN) features under the framework of Bi-LSTM. 

\begin{figure*}[ht]
\begin{center}
\includegraphics[width=1\textwidth]{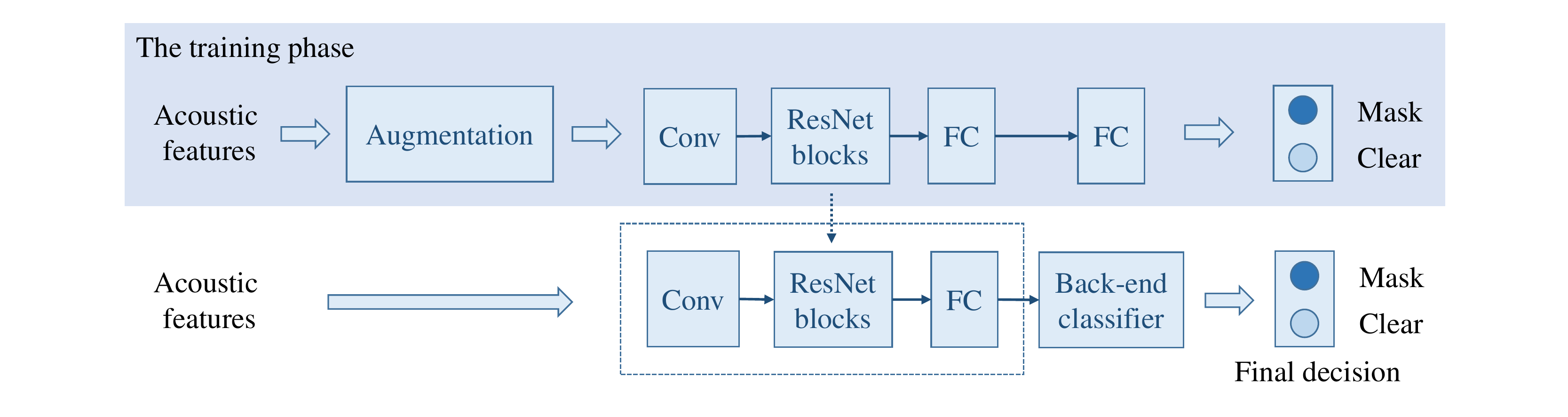}
\end{center}
\vspace*{-0.5cm}
\caption{Framework of the ResNet embedding system with a back-end classifier for the MSC task}
\vspace*{-0.5cm}
\label{emd_framework}
\end{figure*}

\subsection{Feature extraction}

\subsubsection{Gender/Speaker-aware features}



Facial movements vary across genders\cite{Odyssia2009Gender, clark2001three} and people, thus wearing masks may have different effects for different genders and speakers. As mentioned in section \ref{mask_task}, previous studies point out that wearing masks could degrade the speech signal and decrease the performance of speaker recognition \cite{Saeidi2016Analysis}. But whether gender and speaker characteristics could influence mask detection is uncertain. To further explore this question, we introduce the gender-aware and speaker-aware features to the mask detection in this paper.

Motivated by the work of \cite{zhang2018gender} in emotion recognition, we automatically extract gender-aware features from a pre-trained ResNet-based gender classifier. In our work, these features are derived from the penultimate linear layer of the gender classifier network trained with the Voxceleb1 dataset. The ResNet structure is almost the same as described in \cite{cai2020on}, except that the number of output nodes of the penultimate linear layer is 100. To introduce the speaker's information to our training, we follow the configuration of \cite{cai2020on} to train a deep speaker model and extract embeddings as our speaker-aware features. During the optimization of the mask speech classifier, gender, and speaker-aware features are fused on different levels.


\subsubsection{Bottleneck features}

In the BSC, besides raw waveform and log-Fbank features, we investigate bottleneck features as the input of our model. We utilize the BUT/Phonexia bottleneck (BN) feature extractor \cite{Silnova2018, fer2017multilingually} to generate deep phonetic features. 

BN features are extracted from a narrow hidden layer of the acoustic model, whose targets are phonemes. These features contain phonetic information and are suitable for many areas like speech recognition, speaker verification, and language identification. In this toolkit \cite{Silnova2018, fer2017multilingually}, the acoustic model is a stacked bottleneck network with two stages. The first stage is an ordinary bottleneck network, and the second one is built on the bottleneck output of the previous model with a broader context. The BN extractor package provides three trained neural network models, among which we choose the FisherMono and FisherTri models. The FisherMono and FisherTri models are trained on the English Fisher corpus for monophone-states and triphone-states, respectively.

\subsection{Data augmentation}

Data augmentation is a common approach to create corrupted versions of the training data, and hence increase the quantity and diversity of the data. In the MSC task, we adopt three schemes of data augmentation for our training: speed perturbation, SpecAugment\cite{Park2019}, and random erasing, respectively. The effect of different data augmentation approaches is illustrated in Figure \ref{data_augmentation}.

\subsubsection{Speed perturbation}

Speed perturbation is a simple data augmentation approach, which has proven effective in speech recognition, speaker verification, and paralinguistic attribute recognition \cite{Haiwei2019The, Cai2018Exploring}. It can be easily implemented without any additional noise data. Practically, we apply speed perturbation with factor 0.9, 1.0, and 1.1 to augment the data and pool them together for model training. Random cropping and repeated padding on the time axis are computed to maintain the size of input features.

\subsubsection{SpecAugment}

SpecAugment \cite{Park2019} is a simple but effective technique of augmentation. It has been successfully applied in speech and speaker recognition \cite{Park2019, Wang2020Investigation}. It masks the acoustic features partially on the time and frequency domain to train a model that is robust to distorted features. The scheme directly acts on Fbank and is suitable for on-the-fly augmentation. We choose the following deformations to augment the training data:

\begin{enumerate}
    \item Frequency masking is applied on Fbank with a consecutive frequency channel range of $ [ f_0, f_0 + f )$.  $f$ refers to the pre-defined bandwidth chosen from a uniform distribution from 0 to $F$, and $f_0$ is chosen from $ [ 0, v-f)$. $v$ represents the feature dimension.
    \item Time masking is applied on the time axis with a consecutive frame range of $ [ t_0, t_0 + t )$.  $t$ refers to the number of frames to be masked chosen from a uniform distribution from 0 to $T$, and $t_0$ is the beginning frame selected from $ [ 0, u-t)$. $u$ represents the number of frames.
\end{enumerate}

Time warping is also proposed as one of the strategies of SpecAugment. In our cases, we have applied speed perturbation, which is very close to it. Thus we do not consider it in this paper.

\subsubsection{Random erasing}

Random erasing is a data augmentation method proposed by \cite{zhong2020random} in image processing and has successfully applied in object detection, image classification, and person re-identification. Its motivation is close to SpecAugment that both of them mask the features to increase the robustness of models towards deformations. Instead of covering a band on the frequency or time axis, random erasing selects a rectangle region randomly on the features and then replace its value with zero. 

In the training phase, the input acoustic features in a batch are randomly kept unchanged or erased a rectangle region of arbitrary size. The position, width, and the height of the masking rectangle are randomly selected within its pre-defined ranges. And for each batch of training, different versions of corrupted features can be generated, thus enhance the robustness and the generalization ability of models.

\subsection{Modelings}

\subsubsection{Deep embedding system}

Similar to the baseline deep spectrum system, we also extract embedding features from the deep convolutional neural network. Different from the baseline system, which is pre-trained with image corpus, ours is trained directly for the masked or clear targets in an end-to-end manner. In our work, we implement two networks to extract deep representations, including a modified version of ResNet and DenseNet.

The deep ResNet structure is implemented following \cite{Cai2018Exploring, Lin2019LSTM}, which has three main components: a ResNet front-end module, two parallel global pooling, and a two linear layer structure. The ResNet module is composed of a series of residual blocks. The module projects the input Fbank to feature map $F\in\mathbb{R}^{C, H, W}$. Then, the global average pooling (GAP) layer and global standard deviation pooling (GSP) layer are applied on each channel to generate a concatenated 2$C$-dimensional vector:


\begin{scriptsize}
\begin{equation}
    GAP(\bm{F}_c) = \frac{1}{H\times W} \sum_{h=1}^{H} \sum_{w=1}^{W} \bm{F}_{c, h, w},
\end{equation}
\end{scriptsize}

\begin{scriptsize}
\begin{equation}
    GSP(\bm{F}_c) = \frac{1}{H\times W} \sqrt{\sum_{h=1}^{H} \sum_{w=1}^{W} (\bm{F}_{c, h, w} - GAP(\bm{F}_c))^2}.
\end{equation}
\end{scriptsize}

The output vectors of the pooling layer are then fed into the fully-connected layers to make predictions. The embedding features are extracted from the output of the penultimate fully-connected layer.

Our DenseNet \cite{huang2017densenet} structure follows the implementation of torchvision. DenseNet connects every layer with other layers in a feedforward manner, thus has the potential to reduce the problem of gradient vanishing. The deep embedding features are extracted from the output of the average pooling layer.

After training the model, we feed the extracted embeddings to a back-end SVM for predictions. The framework is illustrated in Figure \ref{emd_framework}.

\subsubsection{Bi-LSTM}

The BSC task can be regarded as a sequence-to-sequence regression task. We implement a Bi-LSTM network, a typical sequential modeling method considering speech context from both directions. In our work, two Bi-LSTM layers are stacked, each with 256 units per direction and a dropout rate of 0.6, followed by a fully-connected layer. The final activation function we use is the tanh function.


Through this structure, the input features are transformed into a sequence of predictions. Then, we compute the cosine distance between the ground true upper belt signal as the loss function to update the model.



\section{Experiments and results}

\subsection{Mask Sub-Challenge}

\subsubsection{Setup}

The MSC task is a binary classification task that identifies whether the speaker is wearing a mask or not. The Metric of this task is unweighted average recall (UAR). We investigate different kinds of data augmentation schemes and fusion methods of the gender/speaker-aware feature in our work.

Our augmentation schemes include speed perturbation, SpecAugment, and random erasing. The parameter ($F$, $T$) of SpecAugment is chosen to be (12, 20), which achieves the best performances among the candidates $\{(8, 12), (10, 16), $ $(12, 20), (14, 24), (16, 28)\}$ in our preliminary experiments. In the random erasing method, to explore a suitable proportion range of erased area against input features, we try the candidates $\{(0.02, k)\}, k=0.1, 0.15, 0.2, 0.25, 0.3$, and find out that $(0.02, 0.2)$ is an appropriate choice in our task.

Both the gender-aware and speaker-aware features are extracted from a pre-trained classifier. In our experiments, we introduce these features on two different levels. The first one is the feature level, in which the gender/speaker embeddings are stacked on top of Fbank (Feat-level). Another one is the pooling level. Gender/speaker-aware features are concatenated with the output of the pooling layer (Emb-level). 

Categorical cross-entropy is taken as the loss function. Networks are optimized using stochastic gradient descent (SGD) with Nesterov momentum 0.9 for 100 epochs. During the training process, the learning rate is first initialized as 0.01 and reduced by a factor of 10 when the training loss plateaus. Embeddings are extracted from the trained networks and then fed into an SVM classifier with the BRF kernel with default parameters. We fuse the systems by averaging the output probabilities of different models. 

\subsubsection{Results}


Table \ref{result_mask2} shows the contribution of gender-aware and speaker-aware features under different fusion methods. The performance for gender-aware and speaker-aware features fused in the feature level gives a considerably better result than others, suggesting that both gender information and speaker information are effective for mask detection. It indicates that wearing a mask may bring different effects for different genders and speakers. The UAR score of the embedding level fusion is close to the baseline, which may be caused by the redundancy of gender/speaker information in the embedding layer.

\begin{table}[h]
\centering
\caption{UAR(\%) for gender-aware and speaker-aware feature on the MSC development set}  
\begin{tabular}{@{}cccc@{}}
\toprule
Fusion method & \multicolumn{1}{l}{Fbank(None)} & Feat-level (*) & Emb-level \\ \midrule
Gender        & \multirow{2}{*}{70.7}    & 72.1          & 70.4          \\
Speaker       &                          & 72.2          & 70.7          \\ \bottomrule
\end{tabular}
\label{result_mask2}
\vspace*{-0.2cm}
\end{table}


\begin{table}[h]
    \centering
    \caption{UAR(\%) on the MSC development set and test set}
    \label{result_mask}
    \begin{tabular}{l@{\ }l@{\ \ }c@{\ \ }c}
        \toprule
        Model & Augmentation & Devel & Test\\
        \midrule
        \multirow{5}{*}{ResNet}&(None) & 68.9    & -    \\
        & Speed  & 70.7    & -    \\
        & Speed + SpecAugment (*) & 72.4    & -    \\
        & Speed + RandomErase (*) & 71.5    & -    \\
        & All & 71.6    & -    \\
        \midrule
        \multirow{5}{*}{DenseNet}&(None) & 66.5    & -    \\
        & Speed  & 70.2    & -    \\
        & Speed + SpecAugment (*) & 70.6    & -    \\
        & Speed + RandomErase (*) & 71.2    & -    \\
        & All & 71.2    & -    \\
        \midrule
        The baseline system & - & - & 71.8    \\
        Fused system & - & 73.9 & \textbf{78.8}    \\
        \bottomrule
    \end{tabular}
    \vspace*{-0.2cm}
\end{table}


Experimental results are shown in Table \ref{result_mask}. The ResNet based system achieves a better performance than the DenseNet based system with or without augmentation. 

All the data augmentation methods we apply in the MSC help improve the performance significantly. The SpecAugment approach manages to achieve a more significant growth than the random erasing on the development set. Combining SpecAugment and random erasing does not bring any further improvement, which means that the effect of SpecAugment and random erasing are not complementary.

Our final submitted system is fused with systems having (*) in the Tables \ref{result_mask2} and \ref{result_mask}. Its performance significantly outperforms the baseline system on the test set, which indicates that our approach is robust and effective.


\subsection{Breathing Sub-Challenge}

\subsubsection{Setup}

The BSC task can be viewed as a sequence-to-sequence task. PCC is used as a metric. We train a two-layer stacked Bi-LSTM network with Fbank and BN features to predict the upper belt signal from speech. Models are optimized using Adam for 100 epochs with a batch size of 16. We exploit the BUT/Phonexia bottleneck (BN) feature extractor to extract the 80-dimensional FisherMono and FisherTri features. In the Fbank system, the speech of four minutes is transformed into a 6000-frames acoustic feature sequence with a 60ms window and 40 ms shift. For the BNF system, with a shift of 10ms, we stack four frames together to generate a 6000-frames feature.

\subsubsection{Results}

Comparing the results of the BNF features in Table \ref{table_breathing}, we can find that the BNF system based on triphone-states works slightly better than monophone-states. They achieve higher scores than the baseline system but do not show evident advantages over Fbank. In the test set, our fused system outperforms the baseline system.

\begin{table}[h]
    \centering
    \vspace*{-0.2cm}
    \caption{Results on the BSC development set and test set. The FT. and FM. denote the systems trained with only FisherTri and FisherMono BNF. The baseline system is the end2end framework provided by organizer. The fused system is our submitted system.}
    \label{table_breathing}
    \begin{tabular}{l@{\ \ \ \ }c@{\ \ \ \ }c@{\ \ \ \ }c@{\ \ \ \ }c@{\ \ \ \ }c}
        \toprule
        System & Fbank & FT. & FM. & Baseline & Fused\\
        \midrule
        Devel & 0.530    & 0.536 & 0.530 & 0.507 & 0.545\\
        Test & -    & -    & - &  0.731 & 0.746\\
        \bottomrule
    \end{tabular}
    \vspace*{-0.5cm}
\end{table}


\section{Conclusions}

This paper describes our submitted systems for the MSC and BSC in the Interspeech C{\scriptsize OM}P{\scriptsize AR}E Challenge 2020. For the MSC task, embeddings are extracted from deep convolutional neural networks as representations and fed into a back-end SVM classifier for binary classification. We investigate speed perturbation, SpecAugment, and random erasing as our data augmentation schemes and use the gender/speaker embeddings to further enhance the performance. For the BSC task, we explore Fbank and phonetic features based on the Bi-LSTM structure. Experimental results prove the effectiveness of our data augmentation approaches of the deep embedding systems in the MSC. The performance of the phonetic features is better than the baseline while shows no advantages towards Fbank. Our proposed methods outperform the baselines and achieve 0.746 PCC and 78.8\% UAR on the MSC and BSC evaluation data, respectively.

\section{Acknowledgment}

We want to thank Hamilton, Antonia and Macintyre, Alexis from University College London to share the speech breathing dataset with us for this paper. This research is funded in part by the National Natural Science Foundation of China (61773413), Key Research and Development Program of Jiangsu Province (BE2019054), Six talent peaks project in Jiangsu Province (JY-074), Science and Technology Program of Guangzhou, China (202007030011, 201903010040).

\bibliographystyle{IEEEtran}

\bibliography{mybib}
\vfill\pagebreak

\end{document}